\documentclass[epj,twocolumn]{webofc}
\usepackage[varg]{txfonts}   
\woctitle{Hadron Collider Physics symposium 2012}

\usepackage{lineno}

\usepackage{color}
\usepackage{xspace}
\usepackage{amsmath}
\usepackage{ifthen} 
\newboolean{uprightparticles}
\setboolean{uprightparticles}{false} 

\def\ux85 {\mbox{UX85}\xspace}

\ifthenelse{\boolean{uprightparticles}}%
{

 \def\Pmu         {\ensuremath{\upmu}\xspace}

 \def\PDelta      {\ensuremath{\Delta}\xspace}                 
 \def\PXi      {\ensuremath{\Xi}\xspace}                 
 \def\PLambda      {\ensuremath{\Lambda}\xspace}                 
 \def\PSigma      {\ensuremath{\Sigma}\xspace}                 
 \def\POmega      {\ensuremath{\Omega}\xspace}                 
 \def\PUpsilon      {\ensuremath{\Upsilon}\xspace}                 
 
 \def\PB      {\ensuremath{\mathrm{B}}\xspace}                 
                  
 \def\PD      {\ensuremath{\mathrm{D}}\xspace}

 \def\PK      {\ensuremath{\mathrm{K}}\xspace}

 \def\Pi      {\ensuremath{\mathrm{i}}\xspace}

 \def\Ps      {\ensuremath{\mathrm{s}}\xspace}

}
{

 \def\Pmu         {\ensuremath{\mu}\xspace}

 \mathchardef\PDelta="7101
 \mathchardef\PXi="7104
 \mathchardef\PLambda="7103
 \mathchardef\PSigma="7106
 \mathchardef\POmega="710A
 \mathchardef\PUpsilon="7107
                  
 \def\PB      {\ensuremath{B}\xspace}                 
                  
 \def\PD      {\ensuremath{D}\xspace}

 \def\PK      {\ensuremath{K}\xspace}

 \def\Pi      {\ensuremath{i}\xspace}

 \def\Ps      {\ensuremath{s}\xspace}

}

\def\mup        {\ensuremath{\Pmu^+}\xspace}
\def\mun        {\ensuremath{\Pmu^-}\xspace} 

\def\squark    {\ensuremath{\Ps}\xspace}

\def\kaon  {\ensuremath{\PK}\xspace}
  \def\Kbar  {\kern 0.2em\overline{\kern -0.2em \PK}{}\xspace}

\def\Kz    {\ensuremath{\kaon^0}\xspace}
\def\Kzb   {\ensuremath{\Kbar^0}\xspace}
\def\KzKzb {\ensuremath{\Kz \kern -0.16em \Kzb}\xspace}
\def\Kp    {\ensuremath{\kaon^+}\xspace}
\def\Km    {\ensuremath{\kaon^-}\xspace}

\def\KpKm  {\ensuremath{\Kp \kern -0.16em \Km}\xspace}

  \def\Dbar    {\kern 0.2em\overline{\kern -0.2em \PD}{}\xspace}
\def\D       {\ensuremath{\PD}\xspace}

\def\Dz      {\ensuremath{\D^0}\xspace}
\def\Dzb     {\ensuremath{\Dbar^0}\xspace}
\def\DzDzb   {\ensuremath{\Dz {\kern -0.16em \Dzb}}\xspace}
\def\Dp      {\ensuremath{\D^+}\xspace}
\def\Dm      {\ensuremath{\D^-}\xspace}

\def\DpDm    {\ensuremath{\Dp {\kern -0.16em \Dm}}\xspace}

\def\B       {\ensuremath{\PB}\xspace}
  \def\Bbar    {\kern 0.18em\overline{\kern -0.18em \PB}{}\xspace}

\def\Bd      {\ensuremath{\B^0}\xspace}
\def\Bs      {\ensuremath{\B^0_\squark}\xspace}

  \def\Y#1S{\ensuremath{\PUpsilon{(#1S)}}\xspace}

\def\Lbar {\ensuremath{\kern 0.1em\overline{\kern -0.1em\PLambda}}\xspace}

\newcommand{\decay}[2]{\ensuremath{#1\!\to #2}\xspace}         

\def\to                 {\ensuremath{\rightarrow}\xspace}

\def\AT#1     {\ensuremath{A_{\mathrm{T}}^{#1}}\xspace}           

\def\Bsmm     {\decay{\Bs}{\mup\mun}}
\def\Bdmm     {\decay{\Bd}{\mup\mun}}

\def\C#1      {\ensuremath{\mathcal{C}_{#1}}\xspace}                       
\def\Cp#1     {\ensuremath{\mathcal{C}_{#1}^{'}}\xspace}                    
\def\Ceff#1   {\ensuremath{\mathcal{C}_{#1}^{\mathrm{(eff)}}}\xspace}        
\def\Cpeff#1  {\ensuremath{\mathcal{C}_{#1}^{'\mathrm{(eff)}}}\xspace}       
\def\Ope#1    {\ensuremath{\mathcal{O}_{#1}}\xspace}                       
\def\Opep#1   {\ensuremath{\mathcal{O}_{#1}^{'}}\xspace}                    

\newcommand{\tev}{\ensuremath{\mathrm{\,Te\kern -0.1em V}}\xspace}
\newcommand{\gev}{\ensuremath{\mathrm{\,Ge\kern -0.1em V}}\xspace}
\newcommand{\mev}{\ensuremath{\mathrm{\,Me\kern -0.1em V}}\xspace}
\newcommand{\kev}{\ensuremath{\mathrm{\,ke\kern -0.1em V}}\xspace}
\newcommand{\ev}{\ensuremath{\mathrm{\,e\kern -0.1em V}}\xspace}
\newcommand{\gevc}{\ensuremath{{\mathrm{\,Ge\kern -0.1em V\!/}c}}\xspace}
\newcommand{\mevc}{\ensuremath{{\mathrm{\,Me\kern -0.1em V\!/}c}}\xspace}
\newcommand{\gevcc}{\ensuremath{{\mathrm{\,Ge\kern -0.1em V\!/}c^2}}\xspace}
\newcommand{\gevgevcccc}{\ensuremath{{\mathrm{\,Ge\kern -0.1em V^2\!/}c^4}}\xspace}
\newcommand{\mevcc}{\ensuremath{{\mathrm{\,Me\kern -0.1em V\!/}c^2}}\xspace}

\def\invfb   {\ensuremath{\mbox{\,fb}^{-1}}\xspace}

\def\gsim{{~\raise.15em\hbox{$>$}\kern-.85em
          \lower.35em\hbox{$\sim$}~}\xspace}
\def\lsim{{~\raise.15em\hbox{$<$}\kern-.85em
          \lower.35em\hbox{$\sim$}~}\xspace}

\def\pt         {\mbox{$p_{\rm T}$}\xspace}

\def\tell1  {TELL1\xspace}
\def\ukl1   {UKL1\xspace}

\newcommand{\Bsmumu}{\ensuremath{\Bs\to\mu^+\mu^-}\xspace}
\newcommand{\Bdmumu}{\ensuremath{\Bd\to\mu^+\mu^-}\xspace}

\newcommand{\BdKpi}{\ensuremath{\Bd\to K^+\pi^-}\xspace}
\newcommand{\BsKpi}{\ensuremath{\Bs\to K^+\pi^-}\xspace}

\newcommand{\Bhh}{\ensuremath{B^0_{(s)}\to h^+h'^-}\xspace}

\newcommand{\BuJpsiK}{\ensuremath{B^-\to J/\psi K^-}\xspace}

\newcommand{\Bsdmm}{\ensuremath{\ensuremath{B^0_{s,d}}\to\mu^+\mu^-}\xspace}
\newcommand{\Bsd}{\ensuremath{\ensuremath{B^0_{s,d}}}\xspace}

\newcommand{\BRof}[1]{\ensuremath{{\cal B}(#1)}\xspace}

\begin{document}
\title{New results on the search for \Bsmm from LHCb}
%
%

\author{Johannes Albrecht\inst{1,2}\fnsep\thanks{\email{albrecht@cern.ch}} on behalf of the LHCb collaboration}

\institute{CERN, Geneva, Switzerland \and TU Dortmund, Dortmund, Germany}

\abstract{%
A search for the rare decays \Bsmm and \Bdmm is performed with the
LHCb experiment using 1.1\invfb of data collected at $\sqrt s=8\tev$
and 1.0\invfb of data collected at $\sqrt s=7\tev$. An excess of \Bsmm
candidates with respect to the background expectations is observed
with a statistical significance of 3.5 standard deviations. A
branching fraction of $\BRof \Bsmm =(3.2^{+1.5}_{-1.2}) \times 10^{-9}$
is measured with an unbinned maximum likelihood fit. The measured
branching fraction is in agreement with the expectation from the
Standard Model. The observed number of \Bdmm candidates is consistent
with the background expectation and an upper limit on the branching
fraction of $\BRof\Bdmm < 9.4\times10^{-10}$ is obtained.  
}
\maketitle
\section{Introduction}

One of the most important goals of the LHCb experiment at the LHC is to
search for phenomena that cannot be explained by the Standard Model (SM) of
particle physics.  
Precise measurements of the branching fractions of the two Flavour
Changing Neutral Current (FCNC) decays \Bsmm and \Bdmm 
belong to the most promising of these searches. 
Both decays are strongly suppressed by loop and helicity factors,
making the SM branching fraction small~\cite{Buras:2012ru}
\begin{eqnarray}
\BRof{\Bsmm}&=& (3.23 \pm 0.27) \times 10^{-9}\rm{ ~~and
}\label{eq:isidori}\\
\BRof{\Bdmm} &=& (0.11 \pm 0.01) \times 10^{-9}\,.
\end{eqnarray}
These theoretical predictions are the $CP$-averaged branching fractions. 
As pointed out in Ref.~\cite{deBruyn:2012wk}, the finite width
difference of the \Bs system needs to be considered. The time
integrated branching fraction is evaluated to be 
\begin{equation}
\BRof{\Bsmm}_{SM, \langle t \rangle} = 3.4 \times 10^{-9}\, , \label{eq:smbsmm}
\end{equation}
for which the SM prediction (Eq.~\ref{eq:isidori}) and the LHCb
measurement of the width difference
$\Delta \Gamma_s$~\cite{LHCb-CONF-2012-002} are used.  
This is the expected value which is to be compared with an
experimental measurement.

Enhancements of the branching fractions of these decays are predicted in a
variety of different extensions of the Standard Model, an overview is
given in Ref.~\cite{Bediaga:2012py}. In one popular example, the Minimal 
Supersymmetric Standard Model (MSSM), the enhancement is proportional
to $\tan^6\beta$, where $\tan\beta$ is the ratio of the vacuum
expectation values of the two Higgs fields. For large values of $\tan
\beta$, this search belongs to the most sensitive probes for physics
beyond the SM which can be performed at collider experiments. A
review of the experimental status of the searches for \Bsdmm can be
found in~\cite{Albrecht:2012hp}.

\section{Dataset and analysis strategy}


The measurements~\cite{:2012ct} presented here use data recorded by the
LHCb experiment: 1\invfb, recorded in 2011 at an center of mass energy
of $\sqrt s = 7\tev$, combined with 1.1\invfb of data recorded in 2012
at $\sqrt s =8\tev$. The first part of the dataset has already been
analyzed~\cite{Aaij:2012ac} and was used to produce the lowest
published limit on the decay rate of both decays \Bsmm and \Bdmm.

Two main improvements have been implemented over the previous
analysis: the use of particle identification to select \Bhh decays
which are used to calibrate the geometrical and kinematic variables, and
a refined estimate of the exclusive backgrounds. 
The updated estimate of the exclusive backgrounds is also applied to
the 2011 data and the results reevaluated. The results obtained with
the combined 2011 and 2012 data sets supersede those of
Ref.~\cite{Aaij:2012ac}. 

Candidate \Bsdmm events are required to be selected by a hardware and
a subsequent software trigger~\cite{Aaij:2012me}, dominantly by single
and dimuon lines. 
The first step of the final analysis is a simple selection, which
removes the dominant part of the background and keeps about 60\% of
the reconstructed signal events. 
A second selection step, based on a Boosted Decision Tree (BDT)
reduces 80\% of the remaining background while retaining 92\% of the
signal. More details on the selections are given in Ref.~\cite{:2012ct}

\subsection{Signal discrimination}
\begin{figure}
\centering
\includegraphics[width=0.47\textwidth]{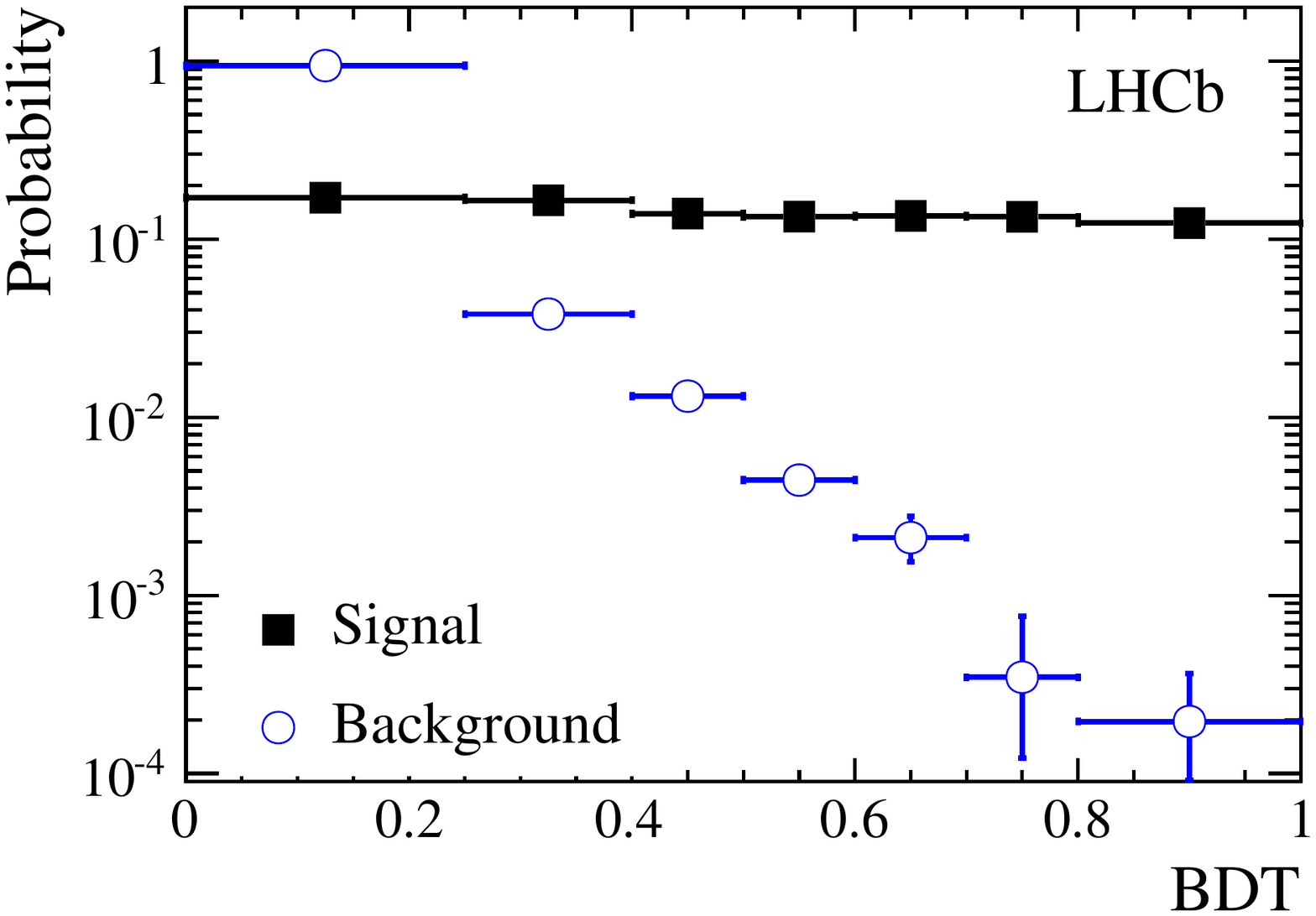}
\caption{BDT distribution for the 2012 dataset, for the signal (black squares)
  and combinatorial background (blue open points). Values normalized to
  the bin size.}
\label{bdt}     
\end{figure}

Each event is then given a probability to be signal or background in a
two-dimensional probability space defined by the dimuon invariant mass
and a multivariate discriminant operator. This likelihood combines 
kinematic and topological variables of the $B^0_{(s)}$ decay using a
BDT. The BDT is defined and trained on
simulated events for both signal and background. The signal BDT shape
is then calibrated using decays of the type $B^0_{(s)} \rightarrow h^+
h^{'-}$, where $h^\pm$ represents a $K^\pm$ or $\pi^\pm$. These
decays have an identical topology to the signal. The calibrated BDT
signal and background shape is shown in Fig.~\ref{bdt}. It is
designed to be flat in the signal, whereas the shape in the
background falls over four orders of magnitude. 

The invariant mass line shape of the signal events is described by a
Crystal Ball function~\cite{crystal}. 
The mass resolution is calibrated with a combination of two methods: an
interpolation of 
$J/\psi$, $\psi(2S)$ and $\Upsilon(1S)$, $\Upsilon(2S)$ and
$\Upsilon(3S)$ decays to two muons and from exclusive \Bhh samples. 
The results are $\sigma_{\Bs}= 25.0 \pm 0.4 $\mevcc and $\sigma_{\Bd}= 24.6
\pm 0.4 $\mevcc. 
The transition point of the radiative tail is obtained from simulated
\Bsmm events smeared to reproduce the mass resolution measured in the
data. 

The background shapes 
are calibrated simultaneously in the mass and the BDT using the
invariant mass sidebands. This procedure ensures that even though the BDT
is defined using simulated events, the result will not be biased by
discrepancies between data and simulation. 

\subsection{Binning}

The binning of the BDT and invariant mass distributions is optimized
using simulation, to maximize the separation between the median of the
test statistic distribution expected for background and SM \Bsmm
signal, and that expected for background only. The chosen number and
size of the bins are a compromise between maximizing the number of
bins and the necessity to have enough \Bhh events to calibrate the
\Bsmm BDT and enough background in the mass sidebands. 


\section{Normalization}

The number of expected signal events is evaluated by normalizing with
channels of known branching fraction. Two independent channels are
used: \BuJpsiK and \BdKpi. The first decay has similar trigger and
muon identification efficiency to the signal but a different number of
particles in the final state, while the second channel has the same
two-body topology but is selected with a hadronic trigger. The event
selection for these channels is specifically designed to be as close
as possible to the signal selection. 
The normalization for \Bsmumu and \Bdmumu is then given as 
\begin{eqnarray}
&&\BRof \Bsdmm   \nonumber\\
&=&{\cal B}_{\rm norm} \times
\frac{\rm \epsilon_{norm}}{\rm \epsilon_{sig}}  \times
\frac{ f_{\rm norm}}{ f_{d(s)}} \times 
\frac{N_{\Bsdmm}}{N_{\rm norm}}  \nonumber \\
& = & \alpha^{\rm norm}_{\Bsd} \times N_{\Bsdmm},
\label{eq:normalization}
\end{eqnarray}
\begin{figure}[pb]
\centering
\includegraphics[width=0.47\textwidth]{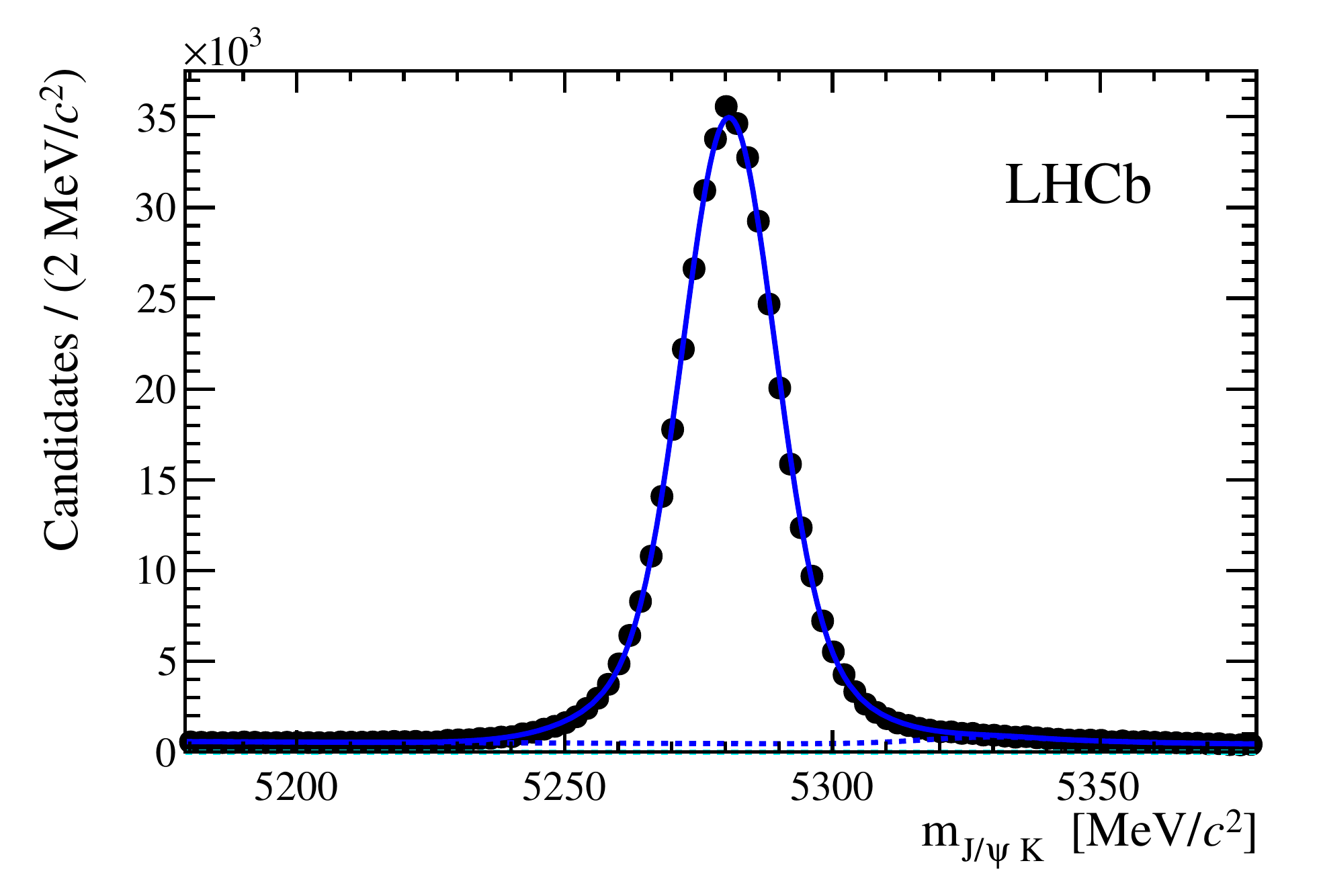}
\includegraphics[width=0.47\textwidth]{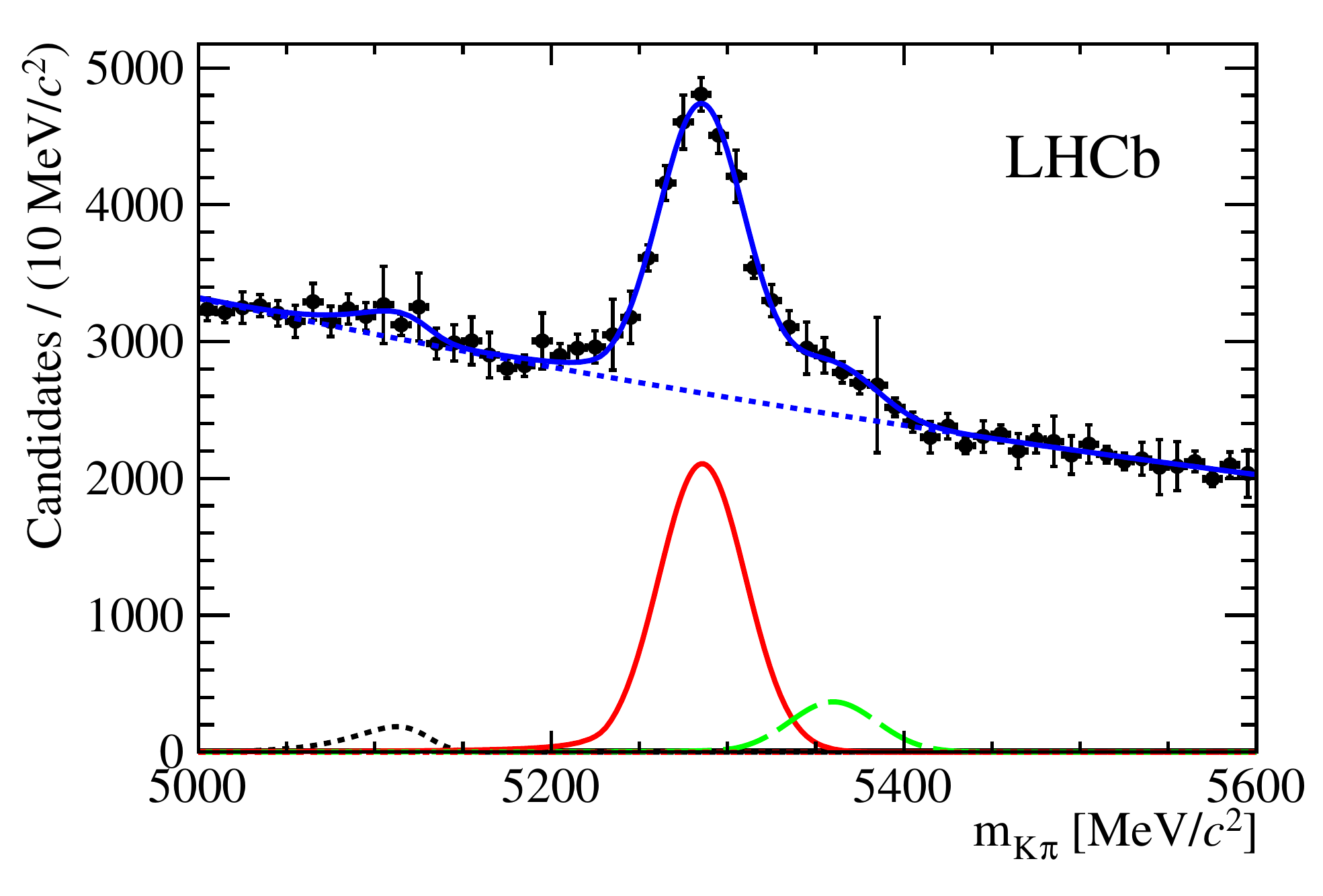}
\caption
{Invariant mass distribution of \BuJpsiK (top) and \BdKpi (bottom)
  candidates. The fit to data is superimposed in blue while the blue
  dotted line is the combinatorial background. In the fit to the
  \BdKpi data, the \BdKpi component (red line), the \BsKpi (green
  dashed line), and the partially reconstructed background (black
  dotted line) are also shown.}
\label{fig:norm}
\end{figure} 
where $f_{d(s)}$ and $f_{\rm norm}$ are the probabilities
that a $b$ quark fragments into a $B^0_{(s)}$ and into the hadron involved
in the given normalization mode respectively.
The recently updated value $f_s/f_d = 0.256 \pm
0.020$~\cite{LHCb-PAPER-2012-037} is used.
${\cal B}_{\rm norm}$ indicates the branching fraction 
and $N_{\rm norm}$ the number of signal events in the normalization 
channel obtained from a fit to the invariant mass distribution.
The efficiency ${\rm \epsilon_{sig(norm)}}$ for the signal (normalization channel) is
the product of the reconstruction efficiency of all the final state particles of the decay 
including the geometric acceptance of the detector, 
the selection efficiency for reconstructed events, and
the trigger efficiency for reconstructed and selected events. 
$N_{\Bsdmm}$ is the number of observed signal events.
The ratios of reconstruction and selection efficiencies are estimated
from the simulation, while the ratios of trigger efficiencies on
selected events are determined from data.

The fit to the two normalization channels is shown in
Fig.~\ref{fig:norm}. The observed numbers of \BuJpsiK and \BdKpi
candidates are $424\,222\pm 1\,452$ and $14\,579\pm 1\,110$.
The two normalization factors are in agreement within the
uncertainties and their weighted average, taking correlations into
account, is
\begin{eqnarray}
\alpha^{\rm norm}_{\Bs}&=& (2.80 \pm 0.25) \times 10^{-10} { \rm ~~and}\nonumber\\
\alpha^{\rm norm}_{\Bd}&=& (7.16 \pm 0.34) \times 10^{-11} \, ,
\end{eqnarray}
for \Bsmumu and \Bdmumu candidates inside a signal window of
$\pm60\mevcc$ around the mass central value. These normalization factors
are used for the limit computation. The normalization factors in the
full mass range, used in the fit for the branching fraction, are 10\%
lower.

\section{Background characterization}
\label{sec:bkg}

Partially reconstructed decays of beauty mesons or baryons can
pollute the low mass sidebands. The dominant modes are:
\begin{itemize}
\item $B^0 \to \pi^- \mu^+ \overline{\nu}$\, ,
\item $B^0 \to \pi^0 \mu^+ \mu^-$ and $B^+ \to \pi^+ \mu^+ \mu^-$\, 
\item \Bhh (with $h^{(')} = K, \pi$)\, and
\item $\Lambda_b \to p \mu^- \overline{\nu}$\, .
\end{itemize}
In some of these modes kaons, pions and protons are misidentified as
muons. 
The contributions of these decays to the \Bsdmm analysis is estimated
from Monte Carlo simulated samples by folding the $K\to\mu$, $\pi \to
\mu$  and $p \to \mu$ fake rates extracted from $D^0 \to K^+ \pi^-$
and $\Lambda \to p \pi^-$  data samples into the spectrum of simulated
events. 
The fractional yields in BDT bins and the parameters that
describe the mass lineshape are used as nuisance
parameters in the unbinned maximum likelihood fit used to determine the
branching fraction.



The exclusive background contribution in the most sensitive region of
the analysis (BDT$>0.8$) is dominated by the 
$B^0 \to \pi^- \mu^+ \overline{\nu}$, \Bhh and 
$B^{0,+} \to \pi^{0,+} \mu^+ \mu^-$ decays.


\section{Results}
%

\begin{figure*}[tp]
\centering
\includegraphics[width=\textwidth]{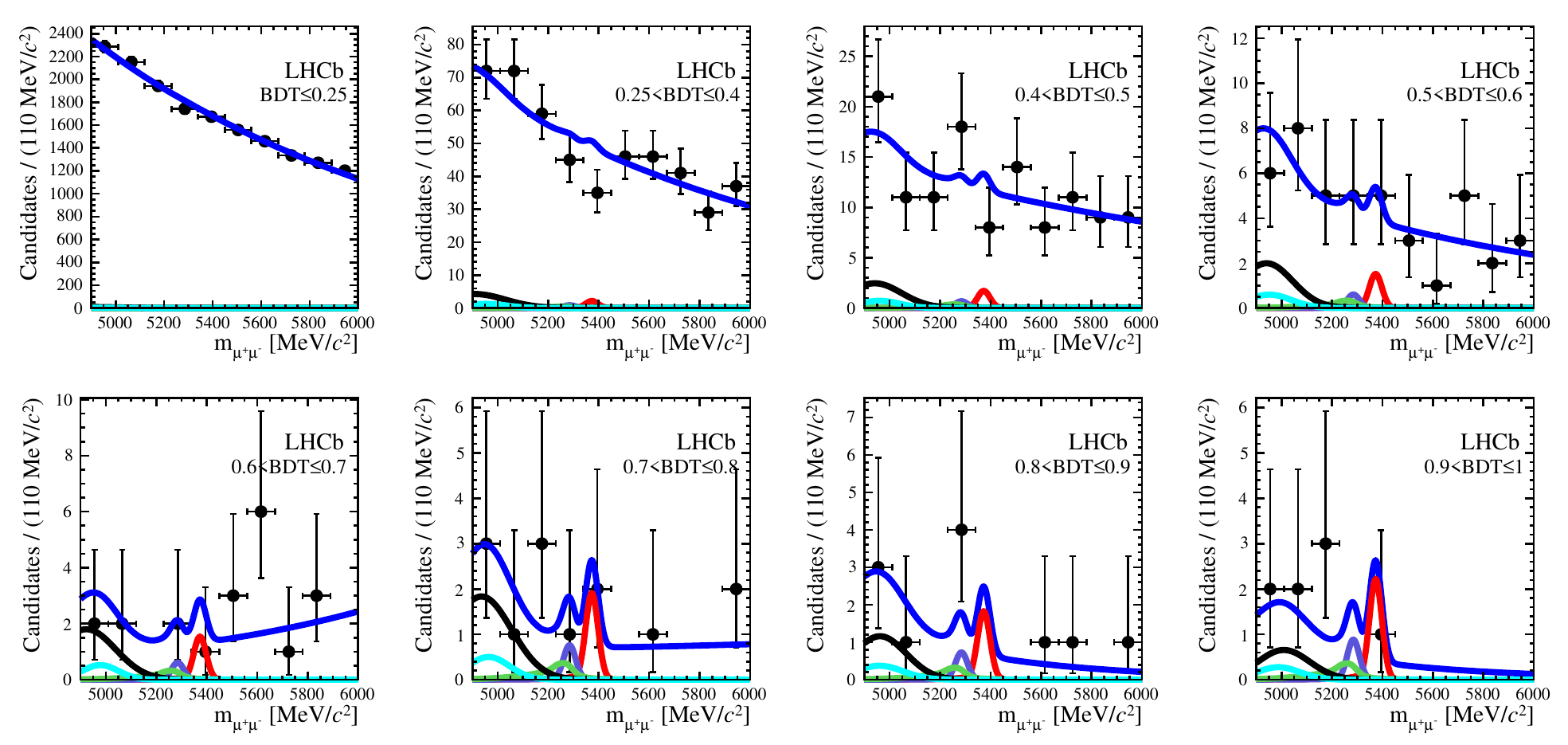}\vspace{5mm}
\includegraphics[width=\textwidth]{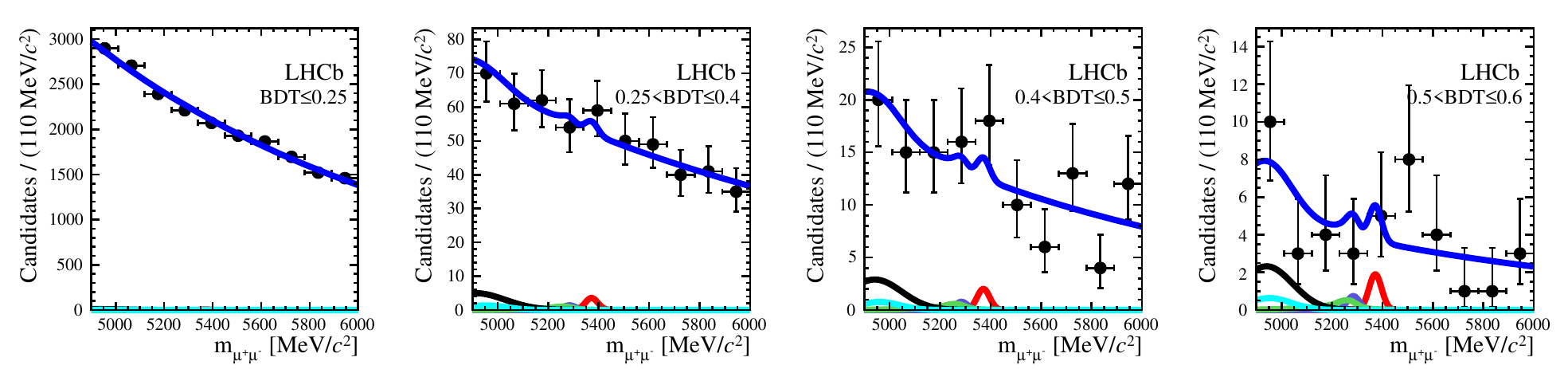}
\includegraphics[width=0.75\textwidth]{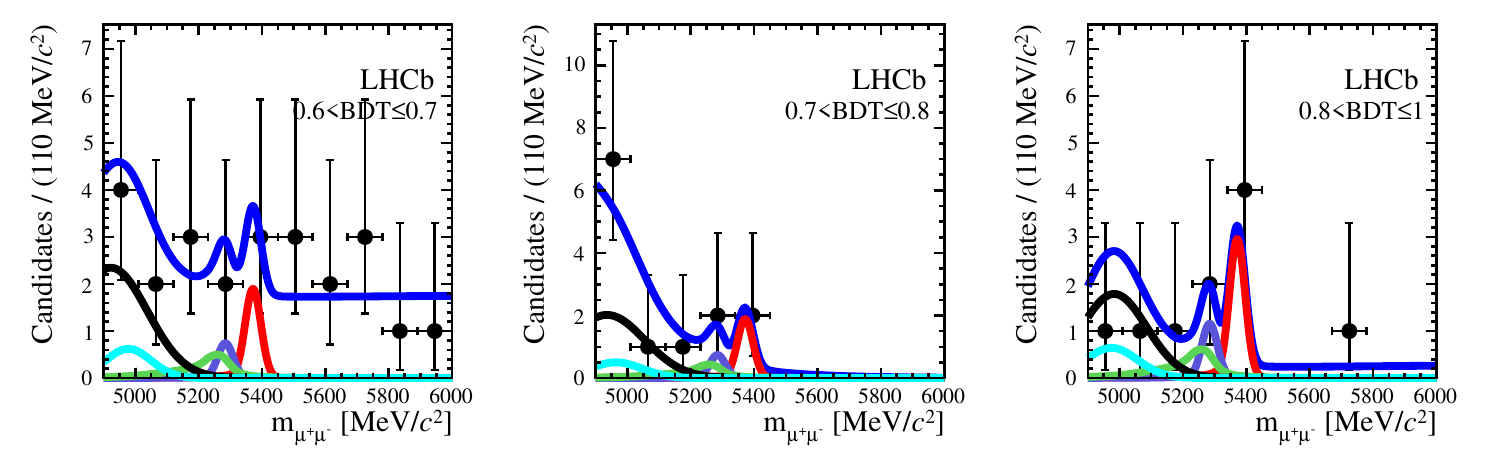}\includegraphics[]{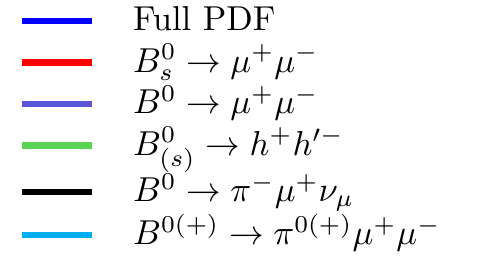}
\caption
{Simultaneous fit of the invariant mass distribution in the 8 BDT bins
  of 2011 (top) and 7 BDT bins of 2012 data (bottom).
  The fit result is superimposed in blue, the individual components
  are given in the legend. }
\label{fig:Bsmm_sig}
\end{figure*} 
The observed pattern of events in the 15 BDT bins (8 in the 2011 data
and 7 in the 2012 data) is shown in Fig.~\ref{fig:Bsmm_sig} for \Bsmm
(top) and \Bdmm (bottom) together with the fit for the branching
fraction, which includes components for the exclusive background
components discussed in Sec.~\ref{sec:bkg}.


%

\begin{figure}[]
\centering
\includegraphics[width=0.47\textwidth]{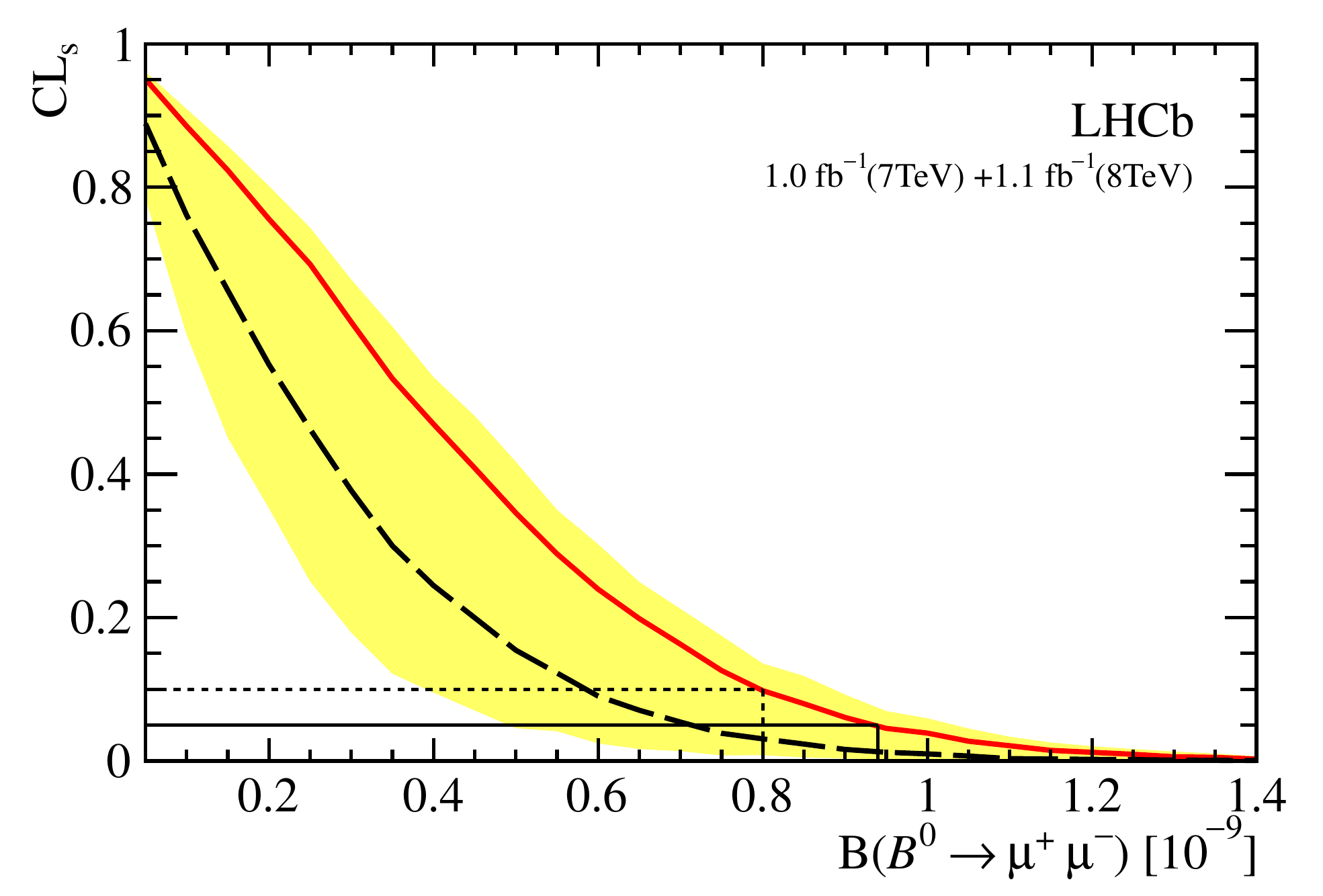}
\includegraphics[width=0.47\textwidth]{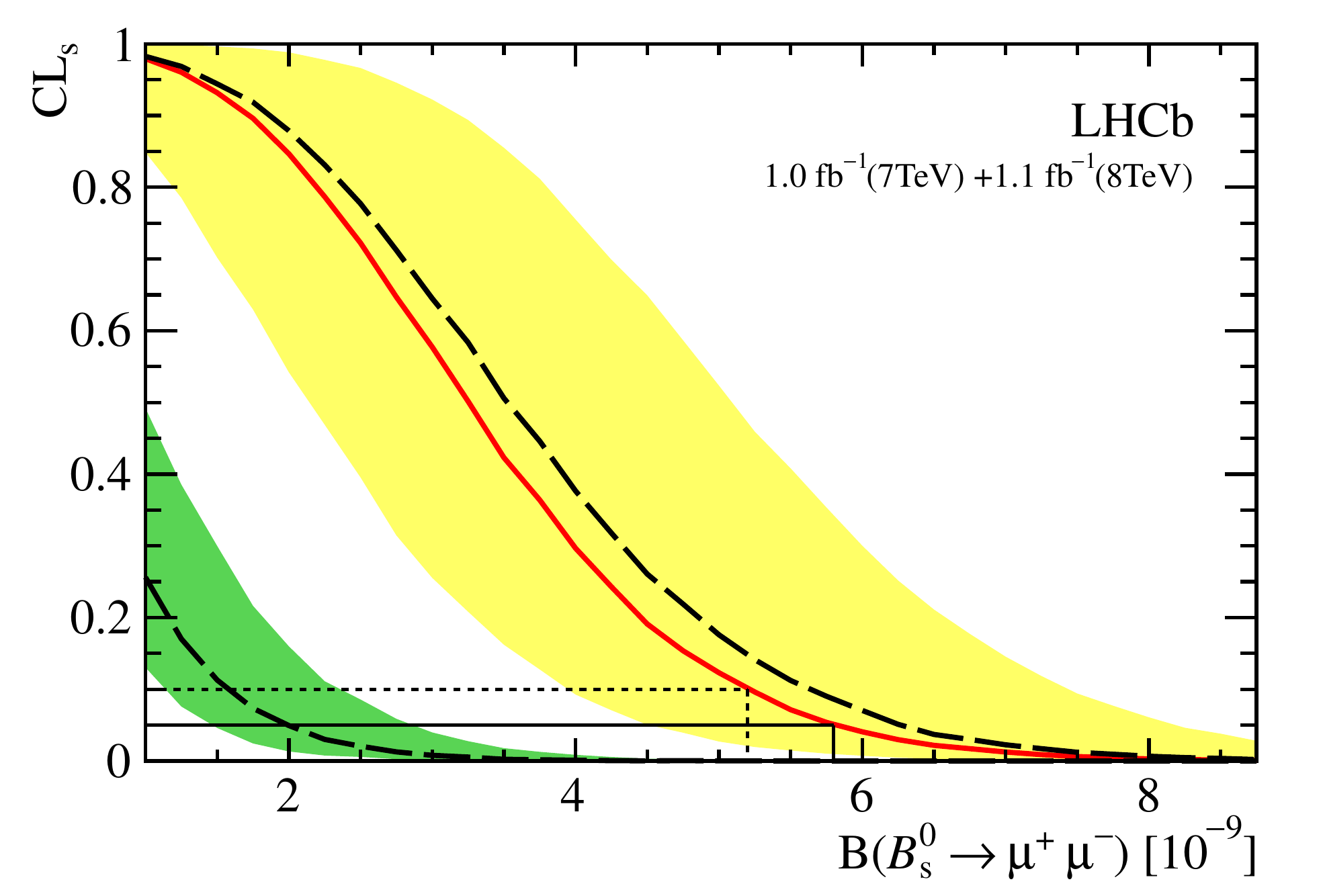}
\caption
{CLs as a function of the assumed \BRof for \Bdmm (top) and \Bsmm
  (bottom) decays for the combined 2011+2012 dataset. The long dashed
  gray curves are the medians of the expected CLs distributions if
  background and SM signal were observed. The 
  yellow area covers the 1$\sigma$ area around the median.
  The solid red curves are the observed CLs. For the \Bsmm (bottom), the
  long dashed gray curve in the green area is the expected CLs
  distribution if background only was observed with its 1$\sigma$
  interval.}
\label{fig:cls}
\end{figure} 

The number of expected combinatorial background events in the \Bd and
\Bs search windows is determined from a simultaneous unbinned
likelihood fit to the mass projections in the BDT bins.  
The same fit is then performed on the full mass range 
to extract the \Bsmumu and \Bdmumu branching fractions.

In this fit the parameters that describe the mass distributions of the exclusive backgrounds, 
their fractional yield in each BDT bin and their overall yields are constrained 
to vary within $\pm 1 \sigma$ with respect to the expected values.
The combinatorial background is parameterized with an exponential function with 
a slope and a normalization which are free parameters of the fit.

The \Bsmumu and \Bdmumu signal yields are free parameters of the fit.  
Their fractional yields in BDT bins are constrained to the BDT fractions calibrated with the
\Bhh sample and the parameters of the Crystal Ball functions 
that describe the mass lineshape are constrained to vary within $\pm 1
\sigma$ with respect to the expected values.

The systematic uncertainties in the exclusive background and signal
predictions in each bin are computed by fluctuating the mass
parameters, the BDT fractional yields and the normalization factors 
along the Gaussian  distributions defined by their associated  uncertainties.
The systematic uncertainty on the estimated number of combinatorial 
background events in the search windows is computed by fluctuating 
with a Poissonian distribution the number of events measured in the sidebands, 
and by varying the value of the exponent accordingly to the its uncertainty.


The compatibility of the observed distribution of events with a given
branching fraction hypothesis is computed using the CLs
method~\cite{Junk:1999kv,0954-3899-28-10-313}. The pattern observed
for \Bdmm decays is compatible with the background only
hypothesis. The CLs curve is shown in Fig.~\ref{fig:cls}~(top). 
The observed CLb value at $CL_{s+b} = 0.5$ is 89\%. 

An excess of \Bsmm candidates is observed, the $CL_s$ curve to
evaluate its significance is shown in Fig.~\ref{fig:cls}~(bottom). 
The probability that background processes can produce the observed
number of \Bsmumu candidates or more is  $5 \times 10^{-4}$
and corresponds to a statistical significance of about 3.5 standard deviations.
The values of the \Bsmumu branching fraction extracted from the fit is
\BRof\Bsmumu = $(3.2^{+1.5}_{-1.2}) \times 10^{-9}$, in good agreement
with the SM prediction.


\pagebreak
\section{Conclusions}

A search for the rare decays \Bsmumu and \Bdmumu has been performed
with 1.1\invfb of data collected at $\sqrt{s}$ = 8\tev and 1.0\invfb
of data collected at $\sqrt{s}$ = 7\tev. 
The data in the \Bd search window are consistent with the background
expectations and an upper limit of \BRof\Bdmumu $< 9.4  \times
10^{-10}$ is obtained at 95\% CL. This is the most stringent published
limit on this decay rate. 
The data in the \Bs search window show an excess of events with
respect to the background expectation with a statistical
significance of  3.5 $\sigma$. 
A branching fraction of  \BRof \Bsmumu $= (3.2^{+1.5}_{-1.2}) \times
10^{-9}$ is measured. This is the first evidence of the \Bsmumu
decay. 

\vskip 2mm
The next step is a precision measurement of the decay rate of \Bsmm
and then to limit and then measure the ratio of the decay rates of
\Bsmm/\Bdmm. 
This ratio allows a stringent test of the hypothesis of minimal flavor
violation and a good discrimination between various extensions of the
Standard Model. 

It should be stated that the precise measurement of \BRof{\Bsdmm}
provides complementary information to the searches performed at high
\pt experiments.

\bibliography{article}


\end{document}